\documentclass[12pt]{article}

\usepackage{amsmath}
\usepackage{amsfonts}
\usepackage{latexsym}

\begin{document}


\thispagestyle{empty}

\begin{flushright}
\begin{tabular}{l}
hep-th/0103219 \\
 MIT-CTP-3102 
\end{tabular}
\end{flushright}

\vspace{10mm}

\begin{center}

{\Large \bf A Field-theoretical Interpretation of the}\\ 
\vskip .12in
{\Large \bf Holographic Renormalization Group}\\
\vspace{10mm}

\vspace{10mm} 

{\bf Johanna Erdmenger} \\[3mm] 
{\em
Center for Theoretical Physics \\ 
Massachusetts Institute of Technology \\
77 Massachusetts Avenue
 \\ 
Cambridge, MA  02139-4307 } \\[3mm] 

{\em email: jke@mitlns.mit.edu}

\vspace{10mm}

{\bf Abstract}  
\end{center}

A quantum-field theoretical interpretation is given to the holographic 
RG equation by relating it to a field-theoretical local RG equation
which determines how Weyl invariance is broken in a quantized field theory.
Using this approach we determine the relation between the holographic
C theorem and the C theorem in two-dimensional quantum field theory
which relies on the Zamolodchikov metric.  Similarly we discuss how 
in four dimensions the holographic C function is related to a
conjectured field-theoretical C function. The scheme
dependence of the holographic RG due to the possible presence 
of finite local counterterms is discussed in detail, as well
as its implications for the holographic C function. 
We also discuss issues special to the situation when mass 
deformations are present. Furthermore we suggest that the holographic
RG equation may also be obtained from a bulk diffeomorphism
which reduces to a Weyl transformation on the boundary.

\medskip


\break


\renewcommand{\theequation}{\thesection.\arabic{equation}}


\def\A{{\cal A}}
\def\B{{\cal B}}
\def\O{{\cal O}}
\def\ga{\gamma}
\def\be{\beta}
\def\si{\sigma}
\def\la{{\lambda}}
\def\O{{\cal O}}

\newcommand{\N}{{\cal N}}

\def\e{{\rm e}}
\def\d{{\rm d}}
\def\s{{\rm s}}
\def\pr{\partial}
\def\D{{\cal D}}
\def\L{{\cal L}}

\renewcommand{\epsilon}{\varepsilon}

\newcommand{\half}{{\textstyle \frac{1}{2}}}
\def\quar{{\textstyle \frac{1}{4}}}

\def\Ga{\Gamma}

\renewcommand{\i}{{\rm i}}

\newcommand{\w}[2]{{\ts {{\hfill #1} \atop {\hfill #2}}\!w}{}}
\newcommand{\wb}[2]{{\ts {{\hfill #1} \atop {\hfill #2}}\!\bar w}{}}

\def\W{{\cal W}}

\newcommand{\al}{\alpha}
\def\l{\langle}
\def\r{\rangle}

\def\de{\delta}
\def\si{\sigma}
\def\ga{\gamma}
\def\la{\lambda}
\def\ka{\kappa}

\newcommand{\eps}{\varepsilon}

\def\ts{\textstyle}

\newcommand{\dx}{\!\!{\rm d}^4x\,\,}

\def\E{{\cal E}}
\def\A{{\cal A}}

\newcommand{\gomn}{g^{\mu\nu}}
\newcommand{\gumn}{g_{\mu\nu}}

\def\achtel{{\ts \frac{1}{8}}}


\newpage 
\setcounter{footnote}{0}

\setcounter{equation}{0}          

\section{Introduction}

In recent years, significant progress has been made towards obtaining
holographic renormalization group (RG) flows from deformations of
AdS space, thereby generalizing the AdS/CFT correspondence to
non-conformal field theories. This line of investigation began 
with the work of \cite{Akhmedov} and \cite{Gomez}.
Subsequently, different approaches have been pursued in order to obtain
the crucial first-order differential equations from the supergravity
equations of motion, in particular using supersymmetry \cite{Freedmanetal}
or the Hamilton-Jacobi equation \cite{Deboer}. These approaches show
in particular
that holographic RG flows are gradient flows governed by a superpotential.
Moreover, a C theorem valid in arbitrary even dimensions has been derived
for deformed AdS spaces \cite{Gomez,Freedmanetal,Girardello,Sahakian}. The
crucial positivity condition arises from the weak energy condition or
equivalently from the Raychaudhuri equation within supergravity.

Within the AdS/CFT correspondence and its generalizations to non-conformal
field theories, the supergravity
fields are sources for composite operators in the dual field theory.
This applies in particular to the supergravity scalars, which are sources
for composite scalar operators. From the quantum 
field theory perspective, we may view these operators as insertions
whose couplings are given by the corresponding sources originating
from supergravity. Therefore the AdS/CFT correspondence and its generalizations
naturally lead us to quantum field theories with space-time
dependent couplings. This implies that also the holographic renormalization
group is local.
For instance the Callan-Symanzik equation derived in \cite{Deboer} 
is of the local form 
\begin{gather} - \, \left(
2 g^{\mu \nu} (x) \frac{\delta}{\delta g^{\mu \nu}(x)} - \beta^i(x)
\frac{\delta}{\delta \phi^i(x)} \right) e^{(S-S_0)}
\hspace{7cm} \nonumber\\ \hspace{4cm} 
 = ({\rm
local \; anomalies \; involving \; the \; curvature \; 
and}\; \pr_\mu \phi^i) \, ,
\label{above}
\end{gather}
with $g^{\mu \nu}(x)$ the metric on a $d$-dimensional hypersurface 
at $r=r_0$. 
The $\beta$ functions in (\ref{above})
are given by derivatives of a superpotential with respect to the couplings.
In a given regularization scheme, the anomalies on the r.h.s.~arise 
from divergent local counterterms $S_0$ necessary to regulate
the supergravity action $S$. Note that the l.h.s.~of 
(\ref{above}) is local - as opposed
to global -  since it involves functional derivatives. 
Of course the l.h.s.~is also non-local in the sense that it contains
operator insertions, and further functional derivatives of the
l.h.s.~will give rise to a Callan-Symanzik equation
 for correlation functions. The divergent counterterms $S_0$ and the
anomaly on the r.h.s.~however are local also in the sense that 
further functional derivatives  give rise to terms involving delta
functions.

The aim of this paper is to give a precise field-theoretical interpretation
for this local holographic flow equation. This local
equation expresses how Weyl symmetry is broken in the dual field theory,
as opposed to the usual `global' Callan-Symanzik equation which
expresses how scale invariance is broken in a quantized field theory.
Within standard quantum field theory,
a local RG equation for general quantum field theories expressing how
Weyl symmetry is broken 
has already been discussed comprehensively by Osborn in
\cite{Osborn}. There, the space-time dependence of the couplings was
used essentially as a trick 
for obtaining a precise definition for insertions
of finite composite operators as functional derivatives of the generating
functional with respect to the couplings.
Furthermore the space-time dependent  couplings allow for the study of Weyl
symmetry consistency conditions for the local anomalies present in the
local RG equation. These have been used in \cite{Osborn} to
give an alternative derivation of the Zamolodchikov C theorem in two
dimensions, using the positivity of the Zamolodchikov metric, which
is defined in terms of the two-point functions of $d=2$-dimensional
operators, to give the
essential positivity of the flow of the C function. Moreover in
\cite{Osborn}, the implications of the 
Weyl consistency conditions were also con\-sidered in four dimensions and
used to relate the RG flow of a conjectured C function to a quadratic
form. However in four dimensions it has so far not been possible to show
that this quadratic form is positive definite.

The essential ingredient for the field-theoretical interpretation of
the holographic RG is to note that the equation (\ref{above}) is precisely of
the form of the field-theoretical RG discussed in \cite{Osborn}.
This identification between the holographic and the field-theoretical
local RG equation is confirmed by calculations of the anomalies on 
the r.h.s.~of 
(\ref{above}) for specific cases within the holographic RG approach 
\cite{Fukuma,Odintsov,Mueck}. These results may all
be expanded in the basis given for these local anomalies within field theory in
\cite{Osborn}.

The identification of the holographic RG with the field theoretical
local RG expressing how Weyl symmetry is broken allows
us to derive a number of new results for the
holographic RG. In particular it gives a field-theoretical
interpretation 	in both two and four dimensions
to the holographic C theorem proved in \cite{Freedmanetal,Girardello}.
  The holographic C function
is given by
\begin{gather} \label{Cfn}
C = \frac{c_0}{|\W(\Phi)|^{d-1}}
\end{gather}
with $\W(\Phi)$ the superpotential generating the flow. 
$c_0$ is a constant chosen such that at the fixed points, $C$ coincides
with the coefficient of the Euler anomaly. The dimension of the bulk
space is $d+1$. The flow of $C$ is positive due to the weak energy
condition. Furthermore it was shown in \cite{Anselmi} that 
the flow of $C$ may also be written in the form
\begin{equation}
 \mu \frac{\d}{\d \mu}{C} = \, - \, {{\ts \frac{2}{3}}} \, 
 C \, L_{ij} \beta^i \beta^j \, \leq 0 \, , \label{Aflow}
\end{equation}
where
$L_{ij}$ is the metric on the space of supergravity scalars.
The result (\ref{Aflow}) is scheme dependent, and the scheme
in which it holds was named `holographic' scheme in \cite{Anselmi}.

Our main result is to show - at least for a particular class of flows -
 that (\ref{Aflow}) coincides with the relation
\begin{gather}
\mu \frac{\d}{\d \mu} {\tilde a} \, = \, 
- \beta^i \pr_i  \tilde{a} = \, - \, \chi_{ij} \beta^i \beta^j \, , \quad
\tilde{a} \equiv a + \beta^i w_i  \, ,\label{Bflow} 
\end{gather}
derived within field theory in \cite{Osborn} using the Weyl consistency
condition. Here $a$ is the coefficient of the Euler anomaly contributing
to the r.h.s.~of (\ref{above}), and $\chi_{ij}$, $w_i$ are the coefficients of
anomaly contributions involving derivatives of the couplings.
Using the results of \cite{Osborn}, we study in detail 
the scheme dependence of $a$, $\chi$, $w$ originating from the 
possibility of adding finite local counterterms to $S_0$ in (\ref{above}).
In two dimensions, the results of \cite{Osborn} allow to relate (\ref{Aflow})
to the field-theoretical C theorem which relies on the Zamolodchikov metric,
but for $d=4$, (\ref{Aflow}) remains a scheme-dependent result from
the quantum field-theoretical perspective.

We calculate $a$, $\chi$, $w$ explicitly for holographic theories
within a minimal subtraction scheme. The essential ingredient for this
calculation is that (\ref{above}) imposes finiteness condition on the
Weyl variation of the divergent local counterterms $S_0$ since the
anomalies on the r.h.s. must be finite. We solve these finiteness
conditions order by order in a strong-coupling perturbative
expansion, in which each order is characterized by the number of derivatives
of the superpotential with respect to the supergravity scalars. Since
we use a cut-off $\eps$ in the radial direction as regulator, this 
procedure corresponds to an extension of the anomaly calculation method
of \cite{Henningson,Skenderis,Petkou}  to scalars with unprotected
dimension. We show that the higher order corrections to
$a$, $\chi$, $w$ cancel. Since our calculation of
$a$, $\chi$, $w$ is in agreement with (\ref{Aflow}), we conclude that the
`holographic' scheme corresponds in fact to minimal subtraction.

Since the holographic RG characterizes
the Weyl transformation properties of holographic
theories, we suggest that it may be possible to give an alternative derivation
of the holographic RG by considering a bulk diffeomorphism which reduces
to a Weyl transformation on the boundary. This will correspond to 
generalizing the so-called PBH transformation \cite{Henneaux}, a bulk
diffeomorphism with this property which has been applied to holographic
theories in \cite{Theisen}, to deformed AdS spaces. As a starting point
for this program we generalize the PBH transformation to scalars in the
conformal case. When conformal symmetry is broken, the scalars
are expected to acquire an anomalous Weyl weight related to their
$\beta$ function.

Moreover we point out that for a complete field-theoretical understanding
of the holographic RG, it will be necessary to show that there is 
agreement between the RG equation and the renormalization procedure
for correlation functions, as there is in standard quantum field theory.
In this context we restrict our attention again to the simple case
when the $\beta$ functions vanish. For this case we show that
there is consistency between the PBH transformation - or
equivalently the holographic RG - at the level of the action and at the level
of the two-point functions. For the general case $\beta^i \neq 0$,
the field-theoretical discussion of \cite{Osborn}
indicates a possibility for proving the agreement between
RG equation and correlator renormalization in general,
and we hope that a generalization of the results presented here
may serve as a basis for applying this analysis in the holographic context.
This would provide  a link between the holographic RG and two-point
functions calculated for deformed AdS spaces for instance in 
\cite{Starinets1,Dewolfe,AFT,Bianchi}.

Our analysis applies to smooth flows from an UV to an IR fixed
point, which may be described within the supergravity approximation.
Typically we consider the case where $\N=4$ supersymmetry at the boundary
is broken down to $\N=2$ or $\N=1$ supersymmetry by operator deformations.
The identification of the holographic RG with the field-theoretical 
local RG and the proposed generalization of the PBH transformation
apply to theories were conformal symmetry is broken either by relevant
(as in the examples given in \cite{Freedmanetal,Girardello}) or by
marginal operator deformations. However the identification of
(\ref{Aflow}) and (\ref{Bflow}) for the holographic C theorem
applies so far only to the case when conformal symmetry is broken
only by marginal operator deformations of engineering dimension $d$. 
It is essential for our analysis that these operators
acquire an anomalous dimension $\gamma_i{}^j = \pr_i \beta^j$ along the flow
\footnote{It will be interesting to construct an explicit example for
such a holographic flow. We leave this for future work.}. 

In the presence of relevant operator deformations, the relation between
(\ref{Aflow}) and (\ref{Bflow}) is less obvious 
due to the following:
Scalars dual to relevant operator deformations contribute anomalies of a
different structure to the r.h.s.~of (\ref{above}) as those dual to marginal
operators. This is due to the fact that the l.h.s.~of (\ref{above})
- and therefore also its r.h.s.~- is of dimension $d$. Since the scalars
$\phi^i$ dual to relevant operators are dimensionful, their 
anomaly contribution to the r.h.s.~of (\ref{above}) will contain less
derivatives than those of the scalars dual to marginal operators.
This coincides with the fact that correlators of lower dimension operators
are less divergent as  distributions. - Of course all of 
the supergravity
scalars $\Phi^i$ appearing in the bulk supergravity action are dimensionless.
However in an expansion in the radial cut-off  $\eps$ of dimension 
(length)$^2$, 
they have the form \cite{Skenderis}
\begin{gather}
\Phi^i(r) = \eps^{(d- \Delta^{(i)} )/2} \, ( \phi^i{}_0(r_0) + 
\phi^i{}_2(r_0) \eps + \dots) \, , \quad \eps = (r-r_0)^2 \, ,
\end{gather}
such that $\phi^i{}_0(r_0)$ is of mass dimension $(d- \Delta^{(i)} )$.
It is $\phi^i{}_0$ and not $\Phi^i$ which contributes to the anomaly in 
(\ref{above}), since (\ref{above}) holds in the case when $\eps=0$.
For flows generated by relevant operators,
we leave a complete study of the field-theory interpretation
of the holographic C theorem for future work. Let us emphasize however
that it was shown in the field theory analysis of \cite{Osborn}
that the relation (\ref{Bflow})  is valid also in the presence
of relevant operators whose coupling is of dimension two, ie.~which
corresponds to a supergravity scalar with $\Delta^{(i)}=2$ in $d=4$.
Still, in the field theory analysis of \cite{Osborn}
 the anomaly coefficients
$a$, $\chi$, $w$ depend only on dimensionless couplings dual to marginal
operators. Therefore when relevant operator deformations are present,
the relation between (\ref{Bflow})
and (\ref{Aflow}) - which involves the $\Phi^i$ - is less clear.

A field-theoretical interpretation of the holographic RG 
complementary to the one presented here was given in \cite{Li}
by relating the holographic RG to the exact renormalization group.
A Wilsonian interpretation of the holographic RG was given in
\cite{Balasubramanian}.

The paper is organized as follows:  In section 2 we give a brief review
of the field-theory results of \cite{Osborn} necessary for our analysis.
In section 3 we discuss the conformal case, for which we generalize the
PBH transformation to scalars and show that there is agreement
between the holographic RG at the level of the action and at the
level of the correlation functions.
We also discuss contributions to the Weyl anomaly for scalars dual
to lower-dimensional relevant operators in this context. In section 4 we 
suggest how to generalize the PBH transformation to deformed AdS spaces.
Furthermore we show that the result (\ref{Aflow}) for the holographic C 
theorem is in agreement with the Weyl consistency condition
for the holographic RG. We discuss its scheme dependence
and use the results of \cite{Osborn}
to relate (\ref{Aflow}) to the field-theoretical C theorem in two dimensions.
In section 5 we calculate the anomaly coefficients of (\ref{Bflow}), which 
are relevant to the C theorem, from finiteness conditions for the Weyl 
variation of the local divergent  counterterms $S_0$ in (\ref{above})
in a strong-coupling perturbative approach.  We conclude by
pointing out further directions for research in section 6.

\section{Local RG equations in QFT}
\setcounter{equation}{0}          

Her we give a brief review of the results for the local RG derived in
\cite{Osborn} which are needed below. For further details we refer to
\cite{Osborn}. We consider a renormalizable quantum field theory in $d$
dimensions which is coupled to a curved space background. Furthermore the
couplings $\lambda_i$ are considered to be space-time dependent. 
We define insertions of the stress tensor and
of scalar composite operators by virtue of 
\begin{equation}
T_{\mu\nu} =  \frac{2}{\sqrt{g}}\frac{\delta S}{\delta
  g^{\mu\nu}}\quad;\quad \O_i =  \, -  \frac{1}{\sqrt{g}}\frac{\delta
  S}{\delta\lambda^i}\quad ,
\end{equation}
where the vacuum energy functional is given by ${\rm exp} W = \int
\D \phi \, {\rm exp} (- S)$.

These definitions ensure in turn a well-defined expression for correlation
functions , eg.~for 
\begin{equation}
\langle \O_i(x) \O_j(y) \rangle = \left.\frac{\delta^2
    W}{\delta\lambda^i(x)\delta\lambda^j(y)}\right|_{\lambda_k =
  const.}\quad,
\end{equation}
and similarly for higher point functions and for the energy-momentum
tensor. The local RG equation discussed in \cite{Osborn} is of the form 
\begin{equation}\label{RG} \, - \, 
\int d^d x\sqrt{g}\sigma(x)\left[ 2 
g^{\mu\nu}\frac{\delta}{\delta g^{\mu\nu}} -
  \beta^i\frac{\delta}{\delta \lambda^i}\right] W = 
\, {{\ts \frac{1}{4 \pi}}} \,
\int \! \d^d x \, \sqrt{g} \left[ \sigma {\cal B} + \pr_\mu \sigma Z^\mu 
\right] \, .
\end{equation}
The right hand side is a local anomaly which involves local terms depending on
the curvature tensors and on derivatives of the coupling. The exact form
of these terms 
depends on the dimension. $\sigma (x)$ is an arbitrary function. 
The interpretation of (\ref{RG}) is that the
terms breaking Weyl symmetry in a renormalizable quantum field theory are
either non-local terms with a $\beta$ function as their coefficient or
local terms. Note that we are only considering dimensionless couplings -
and thus dimension $d$ operators $\O_i$ - in (\ref{RG}).
Mass  terms may be considered separately \cite{Osborn}.

It was shown in \cite{Jack} using a perturbative approach and dimensional
regularisation that examples for both scalar and gauge theories satisfy a
local RG equation of the form (\ref{RG}). Using the BPHZ approach, similar
local RG equations have been derived for $\phi^4$ theory in \cite{Sibold}.

When only dimensionless couplings are present, the vacuum energy 
functional satisfies the relation
\begin{equation} \label{elf}
\left(\mu\frac{\partial}{\partial\mu } + 2\int \d^d x \, 
  g^{\mu\nu}\frac{\delta}{\delta g^{\mu\nu}}\right) W  = 0\quad
\end{equation} 
with $\mu$ the renormalization scale.

In two dimensions the explicit form of the local anomaly terms in
(\ref{RG}) is 
\begin{gather}
 - \, 
\int \ d^2 x \, \sigma \left( 2\, g^{\mu\nu}\frac{\delta }{\delta g^{\mu\nu}}
  - \beta^i \frac{\delta}{\delta\lambda^i}\right) W 
\hspace{6cm} \nonumber\\ \hspace{5cm}
= \, {{\ts \frac{1}{4 \pi}}} \, \int d^2 x \left[
  \sigma \left( \, - a\,  R + \chi_{ij}
    \pr^\mu\lambda^i\pr_\mu\lambda^j\right)
  +\, 2 \, \pr^\mu\sigma w_i \pr_\mu \lambda^i\cdot \right]\quad.
\label{RG2}
\end{gather}

The generating functional $W$ satisfies Wess-Zumino consistency conditions
for Weyl symmetry. Defining 
\begin{equation} \label{Delta}
\Delta [\sigma ] \equiv \,2 \, \int d^d x\sigma g^{\mu\nu }\frac{\delta}{\delta
  g^{\mu\nu} }\quad,\quad \Delta_\beta [\sigma ] \equiv \int d^d
x\sigma\beta^i \frac{\delta}{\delta\lambda^i }\quad,
\end{equation}
the consistency conditions read
\begin{equation} \label{consistency}
\left[ \left(\Delta [\sigma ] - \Delta_\beta [\sigma ]\right),\; \left(
    \Delta [ \sigma' ] - \Delta_\beta [ \sigma ' ]\right)\right] W =
0\quad.
\end{equation}

These conditions imply relations for the coefficients of the local anomaly
terms.
In particular they imply, with $\delta g^{\mu \nu} = 2 \sigma g^{\mu \nu} $,
 $ \delta R = 2 \sigma R + 2 \nabla^2 \sigma$,  
\begin{equation} \label{res2}
\beta^i \partial_i \tilde{a} = \chi_{ij}\beta^i\beta^j \quad;\quad
\tilde{a}\equiv a + \beta^i w_i\quad.
\end{equation}
This relation is invariant under the addition of finite local counterterms
to W. For instance, adding
\begin{gather} \label{sc}
\delta W = \, - \, {{\ts \frac{1}{4 \pi}}} \, \int \! d^2 x \, \sqrt{g}
 \, ( b \, R \, - c_{ij} \pr_\mu \lambda^i \pr^\mu \lambda^j ) \, ,
\end{gather}
to $W$, we have
\begin{gather}
\delta a = \beta^i \pr_i  b \, ,  \; \delta \chi_{ij} = \L_\beta c_{ij} \, , \;
 \delta w_i = - \pr_i b + c_{ij} \beta^j \, , \;  \delta \tilde{a}
= c_{ij} \beta^i \beta^j \, ,
\end{gather}
with $\L_\beta$ the Lie derivative
\begin{gather} \label{Lie}
\L_{\beta} c_{ij} = \beta^k\partial_k c_{ij}
+ \partial_i\beta^k c_{kj} + \partial_j\beta^k c_{ik} \, .
\end{gather}
Differentiating (\ref{RG2}) further with respect to $\lambda_j$ for
$\sigma \equiv 1$ gives
\begin{equation}
\D \langle \O_i (x) \O_j(0) \rangle + \partial_i\beta^k 
\langle \O_k(x) \O_j(0) \rangle + \partial_j
\beta^k \langle \O_i(x) \O_k(0) \rangle = 8\pi \chi_{ij} \partial^2
\delta^{(2)}(x),\label{RGO}
\end{equation}
\begin{equation}
\D \equiv \mu \frac{\partial}{\partial \mu} + \beta^i
\frac{\partial}{\partial \lambda^i }\quad.
\end{equation}

The form of the scalar two-point function is determined by rotational
invariance and the dimension of the operators even when conformal symmetry
is broken. Furthermore it has to be ensured that the two-point function has 
a well-defined Fourier transform such that it does not contain any poles in 
particular for $x \to 0$. A form for the two-point function which meets these 
requirements is given by 
\begin{equation} \label{twopoint}
\langle \O_i (x) \O_j(0)\rangle  
=\partial^2 \partial^2 \Omega_{ij}(t)\;,\quad t =
{\ts \frac{1}{2}} \ln \mu^2 x^2\quad.
\end{equation}
The extraction of derivatives is similar to the method of differential
regularisation \cite{Latorre}.
For the energy-momentum tensor two-point function we have similarly
\begin{gather} \label{pp}
\l T_{\mu \nu} (x) T_{\sigma \rho} (0) \r \, = \, 
( \delta_{\mu \nu} \pr^2 - \pr_\mu \pr_\nu)
( \delta_{\sigma \rho} \pr^2 - \pr_\sigma \pr_\rho) \, \Omega (t) \, .
\end{gather}
Inserting (\ref{twopoint}) into (\ref{RGO})
    we obtain, using  $\partial_\mu(x_\mu / x^2)= 2\pi \delta^{(2)}(x)$,
\begin{equation}
\D\Omega_{ij}'+\partial_i \beta^k\Omega_{kj}' +
\partial_j\beta^k\Omega_{ik}' = 4 \chi_{ij}\quad , \label{D}
\end{equation}
and in an analogous calculation a similar result for $\Omega$ defined
in (\ref{pp}).
The positive definite Zamolodchikov metric is defined by 
\begin{equation}
G_{ij}(x)\equiv {\ts \frac{1}{16}}\,
(x^2)^2 \langle \O_i(x) \O_j(0)\rangle\quad.
\end{equation}
Using (\ref{RGO}) and (\ref{D}) we find
\begin{align} \label{rel2}
G_{ij} & = \,  \quar \,(\Omega_{ij}'' - \Omega_{ij}'''+
{\ts \frac{1}{4}} \Omega_{ij}'''') = \chi_{ij} + \L_\beta c_{ij}\quad,\\
c_{ij} & = \quar (- \Omega_{ij}' +\Omega_{ij}''
-{\ts \frac{1}{4}}\Omega_{ij}''') \, .
\end{align}
The two-dimensional C theorem is then easily obtained by defining the C
function to be 
\begin{equation} \label{ff}
C \equiv 3
 (\tilde{a} + c_{ij} \beta^i \beta^j ) = - \, {\ts \frac{3}{4}} (
\Omega'+ \Omega''-  {\ts \frac{1}{4}} \Omega''')
\end{equation}
which satisfies
\begin{equation} 
C' = - \beta^i\partial_i C = - 3 G_{ij}\beta^i\beta^j \leq 0\quad.
\end{equation}
Note that the linear combination of derivatives of $\Omega$
in (\ref{ff}) corresponds exactly to the C function originally considered by
Zamolodchikov \cite{Zamolodchikov}.

In four dimensions the structure of the local anomaly is more complicated
due to the larger number of possible independent forms. With the 
definitions in (\ref{Delta}) we have
\begin{gather}
\label{RG4} \, - \, 
( \Delta[\sigma] - \Delta_\beta[\sigma]) \, W  = \int\!\! \d^4 x \, 
\sqrt{g} \left[ \sigma {\cal B} + \pr^\mu \sigma Z_\mu \right] \, ,
\end{gather}
where
\begin{align} \label{ano4}
{\cal B} = & \; \;\;c F - a G + {\ts \frac{1}{9}}  \beta_c R^2 \nonumber\\
& + {\ts \frac{1}{3}} \chi^e_i \pr_\mu \lambda^i \pr^\mu R +
{\ts \frac{1}{6}} \chi^f_{ij} \pr_\mu \lambda^i \pr^\mu \la^j
R - \half \chi^g_{ij} \pr_\mu \la^i \pr_\nu \la^j G^{\mu \nu}
+ \half \chi^a_{ij} \nabla^2 \la^i \nabla^2 \la^j 
\nonumber\\ & + \half \chi^b_{ijk}
\pr_\mu \la^i \pr^\mu \la^j \nabla^2 \la^k + {\ts \frac{1}{4}} \chi^c_{ijkl}
\pr_\mu \la^i \pr^\mu \la^j \pr_\nu \la^k \pr^\nu \la^l \, .
\end{align}
Here $R$ is the Ricci scalar, F the square of the Weyl tensor 
$C_{\mu \si \rho \nu}$, $G$ the Euler density and $G_{\mu \nu}$ the 
Einstein tensor. For $Z^\mu$ we have
\begin{align}
Z_\mu = &-G_{\mu \nu} w_i \pr^\nu \la^i + { \ts \frac{1}{3}}
 \pr_\mu ( q R) + {\ts \frac{1}{3}} R Y_i \pr_\mu \lambda^i \nonumber \\
& + \pr_\mu (U_i \nabla^2 \lambda^i + \half V_{ij} \pr_\nu \la^i \pr^\nu
\la^j ) 
+ S_{ij} \pr_\mu \lambda^i \nabla^2 \lambda^j + \half
T_{ijk} \pr_\nu \la^i \pr^\nu \la^j \pr_\mu \la^k \, .
\end{align}
In this case the consistency condition (\ref{consistency}) implies a 
number of relations for the anomaly coefficients, among which
\begin{gather}
\beta^i \pr_i \tilde{a} = {\ts \frac{1}{8}} \chi^g_{ij} \beta^i \beta^j \, ,
\;\;\;\; \tilde{a} \equiv a + {\ts \frac{1}{8}} w_i \beta^i \, .
\label{rel4}
\end{gather}
This is analogous to the relation (\ref{res2}). However in four dimensions
it has not been possible so far to relate $\chi^g_{ij}$ to the 
Zamolodchikov metric, unlike the two-dimensional result (\ref{rel2}).
This may be traced back to the fact that (\ref{RG4}) relates $\chi^g_{ij}$
to the three-point function $\l T_{\mu \nu} \O_i \O_j \r $ rather than
just to the two-point function $\l  \O_i \O_j \r $.

\section{The \boldmath$\beta^i = 0$\unboldmath\ limit of the holographic RG}
\setcounter{equation}{0}          

To show that the holographic RG is consistent with the results presented
in section 2, we begin by considering the simple case in which the $\beta$
functions vanish. We extend the PBH transformation to scalars and show
how it gives rise to a holographic local RG. We discuss the anomalies
present in this RG, and demonstrate consistency with the local RG
for correlation functions. We also discuss operators of lower dimension.

We consider $(d+1)$-dimensional 
AdS space with metric of Fefferman-Graham form
\cite{Fefferman} 
\begin{equation}
  ds^2 = G_{MN}dX^M dX^N = \frac{L^2}{4} \left[
\left(\frac{d\rho}{\rho}\right)^2 +
  \frac{1}{\rho} g_{\mu\nu}(x,\rho)dx^{\mu}dx^{\nu} \right]
\end{equation}
where $M,N = 1,\ldots,5$ and $\mu,\nu = 1,\ldots,4$. The boundary is at
$\rho = 0$. $L$ is the AdS radius,
and $\rho$ is of dimension (length)$^2$. 
Moreover we consider the bosonic part of the supergravity
action\footnote{Throughout we use the conventions of
\cite{Skenderis}.}
\begin{gather} \label{ac}
S = \, \frac{1}{16 \pi G_N} \, \left[ \, 
\int \! d^{(d+1)} x \, \sqrt{G} \left[ R + 2 \Lambda  + 
L_{ij} \pr^\mu \Phi^i \pr_\mu \Phi^j \right]\, - \, \int \! \d^d x \, 
\sqrt{\gamma} \, 2 K \,  \right] \; ,
\end{gather}
where $L_{ij}$ is a metric on the space of supergravity scalars.

As discussed in \cite{Theisen}, there is a $(d+1)$-dimensional diffeomorphism
$v_M$ 
\cite{Henneaux} which reduces to a Weyl transformation on the boundary. This 
is the so-called PBH transformation. With the ansatz
\begin{align}
  \rho = & \rho'e^{2\sigma(x')} \simeq \rho'(1+2\sigma(x'))\quad,\\
  x_\mu =& x'_{\mu} + a_{\mu}(x',\rho')\quad, \qquad v_M=(2\sigma , a_\mu)\,
, 
\end{align}
where the requirement of diffeomorphism invariance of the $(d+1)$-dimensional
metric constrains the form of $a_{\mu}$ , $g^{\mu\nu}$ has the
transformation property
\begin{equation}
  \delta g^{\mu\nu}  \, = \, 
2\sigma(1-\rho \partial \rho)g^{\mu\nu}(x,\rho)
  +\nabla^{\mu}a^{\nu}(x,\rho) + \nabla^{\nu}a^{\mu}(x,\rho)\quad.
\end{equation}
Similarly 
the PBH transformation for the scalars is obtained using
\begin{equation} \label{qq}
\Phi^i(x,\rho) = \, \rho^{(d-\Delta^{(i)})/2} \, \phi^i(x,\rho) \, .
\end{equation}
This gives
\begin{equation}
  \delta \phi^i(x,\rho)  \, = \, 
- \sigma (\Delta^{(i)} - d )\, \phi^i (x,\rho) \,
+ \,2  \sigma\,  \rho \partial \rho  \phi^i
  (x,\rho) \, + \, a^\mu \pr_\mu \phi^i (x,\rho)\quad.
\end{equation}
$i$ is an index in the field space.

For $G^{MN}(x,\rho)\,,\, \Phi^i(x,\rho)$ which are solutions of an
equation of motion, we have the expansion 
\begin{equation}
  \gomn (x,\rho)= \sum_{n=0}^\infty \gomn_{(n)} (x) \rho^n\quad,\quad
  \phi^i(x,\rho) = \, \sum_{n=0}^\infty
  \phi^{i}{}_{(n)}(x)\rho^n\quad.
\end{equation}
For $\Delta^{(i)} = d/2$ we have, instead of (\ref{qq}), 
\cite{Klebanov,Rivelles,Skenderis}
\begin{equation} \label{phitwo}
 \Phi^i(x,\rho) = \rho \ln \rho \sum_{n=0}^\infty
  \phi^{i}{}_{(n)}(x)\rho^n\quad.
\end{equation}

$\phi^{i}{}_{(0)}$ is of mass dimension $d-\Delta^{(i)}$.
At the boundary the PBH transformation reduces to a Weyl transformation
under which in particular
\begin{equation}
  \delta \gomn_{(0)}(x) = 2\sigma\gomn_{(0)}(x)\quad,\quad \delta
  \phi^{i}{}_{(0)}(x)=-(\Delta^{(i)}-d) \sigma \phi^{i}{}_{(0)}(x)\quad.
\end{equation}
As is well known, the action (\ref{ac}) is divergent near the boundary an
requires regularization by subtraction of appropriate divergent
counterterms $S_0$. Introducing a cut-off at $\rho=\epsilon$, the PBH
transformation of the action reduces near the boundary to
\begin{align}
 \, - \, 
\left( 2\epsilon\frac{\delta}{\delta \epsilon} + 2 \gomn_{(0)}\frac{\delta}{\delta
      \gomn_{(0)}} - (\Delta^{(i)} -d
    )\phi^{i}{}_{(0)}
\frac{\delta}{\delta\phi^{i}{}_{(0)}}\right) e^{S-S_0} = \A_4& 
  \label{tr2}\quad.
\end{align}
In (\ref{tr2}) we may take the limit $\epsilon \to 0$ such as to obtain
\begin{equation}
  \label{RGgravity}
\, - \,   \left(2 \gomn_{(0)}\frac{\delta}{\delta\gomn_{(0)}} - (\Delta^{(i)}
 -d
    )\phi^{i}{}_{(0)}\frac{\delta}{\delta\phi^i{}_{(0)}}\right) e^W = \A_4 
\end{equation}
with $e^W \equiv \lim_{\epsilon\to 0} e^{S-S_0}$ the generating functional
and $\A_4$ a local anomaly. This equation coincides with the
field-theoretical local RG equation (\ref{RG}) for the case when
$\beta^i =0$.

We proceed by calculating the local anomaly terms on the r.h.s.\ of
(\ref{RGgravity}) explicitly. We begin with the contributions from
supergravity scalars dual to operators of dimension $d=4$.
As discussed in \cite{Tseytlin} and later in \cite{Minic}, the counterterm
$S_0^{(4)}$
necessary to regulate $S$ as far as fields dual to dimension $d=4$
operators are concerned is given by the action of four-dimensional
conformal supergravity,
\begin{align}
S_0^{(4)} &= - \, \half \, \ln \epsilon \frac{N^2}{4(4\pi^2)}
S_{\rm CSG}\, , \quad S_{\rm CSG} = \int d^4x\sqrt{g}L_{\rm CSG}\,,
\nonumber\\ \label{4dSUGRA}
  L_{\rm CSG} &= 
C^{\mu \nu \rho \sigma} C_{\mu \nu \rho \sigma}
  - \eps_{\mu \nu \al \be} \eps_{\rho \si \ga \de} R^{\mu \nu \rho \si}
  R^{\al \be \ga \de} + 
 \, L_{ij}^{(4)} \phi_{(0)}^i \Delta_4 \phi_{(0)}^j \, ,
\end{align}
where $L_{ij}^{(4)}$ is the metric on the space of supergravity scalars dual
to dimension 4 operators and $\Delta_4$ is the Riegert operator
\cite{Riegert}
\begin{gather}
  \Delta_4 \equiv \nabla^2 \nabla^2 + 2 \nabla^\mu \, (R_{\mu \nu} - 
  {\ts \frac{1}{3}} g_{\mu \nu} R) \, \nabla^\nu \, .
\end{gather}
This result for $S_0^{(4)}$ agrees with the anomaly calculation of
Henningson and Skenderis \cite{Henningson}, extended to scalar fields in
\cite{Skenderis}. Inserting (\ref{4dSUGRA}) into the local RG equation
(\ref{tr2}) with $\Delta = d =4$, and subsequently taking the
limit $\eps \rightarrow 0$,  we obtain for the local anomaly
\begin{equation} \label{a4}
  \A_4 = \frac{N^2}{4(4\pi^2)} L_{\rm CSG}\quad.
\end{equation}
Comparing with the field-theoretical local RG (\ref{RG4}) with
(\ref{ano4}) we identify 
\begin{equation}
  \chi^{a}_{ij} = \frac{N^2}{2(4\pi^2)} L_{ij}^{(4)}\quad,\quad \chi_{ij}^g
  = \frac{N^2}{(4\pi)^2} L_{ij}^{(4)}\quad,\quad c = a =
  \frac{N^2}{4(4\pi)^2}\quad.
\end{equation}
The result for $\chi_{ij}^a$ may also be obtained from considering the
implications of the local RG for the two-point function $\langle \O_i (x)
\O_j(0)\rangle$, which provides a consistency check.
With the definition
\begin{gather} \label{corr}
  \left.\l \O_i(x) \O_j(0) \r \equiv \frac{\delta}{\delta 
\phi^i{}_{(0)}} \frac{\delta}
  {\delta \phi^j{}_{(0)}} {\rm e}^{S-S_0} \, \right|_{\phi^k{}_{(0)}
 = {\rm const}} \, ,
\end{gather}
the standard AdS/CFT procedure gives the result
\begin{equation}
  \langle \O_i(x) \O_j(0)\rangle =
  \frac{N^2}{2 (4\pi)^2}\, L_{ij}^{(4)} \, 
  \partial^2\partial^2\left(\frac{1}{(x^2)^2}- \, \half \, 
\ln\epsilon \, \delta^{(4)}(x)\right)\;,  \label{4corr}
\end{equation}
where the local term guarantees a well-defined expression with a regular
Fourier transform also when $\vec{x} \to 0$. In (\ref{4corr}) ,
derivatives have been extracted following the prescription of differential
regularization \cite{Latorre} in order to obtain a less singular
expression. A similar structure of the AdS/CFT correlators
was discussed in \cite{Chalmers} using
dimensional regularization. 
      
On the other hand, integrating the local RG (\ref{tr2}) for
$\Delta^{(i)}=d=4$ using 
\begin{equation}
  \left(\mu \frac{\pr}{\partial \mu}
 + 2\int d^4x\sqrt{g}\gomn\frac{\delta}{\delta
      \gomn}\right) e^W = 0\;,\; \eps \frac{\pr}{\partial\epsilon} = \int
  d^4x\sqrt{g}\, \eps \, \frac{\delta}{\delta\epsilon}\;,
\end{equation}
and varying the result twice with respect to $\phi^i{}_{(0)}$, we obtain
\begin{equation}\label{RGint}
  \left( \, - \, 2\epsilon \frac{\pr}{\partial\epsilon} +
  \mu \frac{\partial}{\pr \mu} \right) \langle
  \O_i^{(4)}(x)\O_j^{(4)}(0)\rangle = \frac{N^2}{2(4 \pi)^2}\, 
 L_{ij}^{(4)} \partial^2\partial^2
  \delta^{4}(x)\quad.
\end{equation}
It is easy to check that the expression (\ref{4corr}), which is
independent of the renormalizaion scale $\mu$, satisfies (\ref{RGint}).
From the results of \cite{EO} for general conformal field theories,
we expect a similar consistency between the RG equation 
for the energy-momentum tensor two-point function, and the Weyl tensor
squared anomaly in the RG equation at the level of the generating
functional.

For scalar 
operators of dimension $d-2=2$ , the dual scalar $\phi^i{}_{(0)}$
as given by (\ref{phitwo}) is of mass dimension $2$. 
The appropriate boundary counterterm
for regulating the action $S$ is given by 
\begin{equation} \label{S02}
  S_0^{(2)} =    \frac{N^2}{2\pi^2}\, 
\half \ln\epsilon \int d^4x\sqrt{g} L_{ij}^{(2)}\phi^i_{(0)}
\phi^j{}_{(0)}\quad . \quad
\end{equation}
In the RG equation (\ref{tr2}), 
the counterterm (\ref{S02}) gives rise to an anomaly
\begin{gather} \label{a2}
\A_4{}^{(2)} =  \frac{N^2}{2 \pi^2}L_{ij}^{(2)}\phi^i_{(0)}
\phi^j{}_{(0)} \,  ,
\end{gather}
and for the correlation function as defined in (\ref{corr}) we
obtain
\begin{gather}
\l \O_i(x) \O_j (0) \r = \frac{N^2}{2\pi^2}\,  L^{(2)}_{ij} \left(
\frac{1}{(x^2)^2} - \, \half \, \ln \! \eps \, \delta^{(4)} (x) \right) \, ,
\end{gather}
which is consistent with
\begin{equation}\label{RGint2}
  \left( \, - \, 2\epsilon\frac{\pr}{\partial\epsilon} + \mu\frac{\partial}{\pr\mu}
\right) \langle
  \O_i^{(2)}(x)\O_j^{(2)}(0)\rangle = \frac{N^2}{2 \pi^2} \,
  L_{ij}^{(2)} 
  \delta^{4}(x)\quad.
\end{equation}
Note the different structure of the anomalies for operators
with $\Delta^{(i)} = 4$ in (\ref{a4}) with (\ref{4dSUGRA}) and with
$\Delta^{(i)} = 2$ in (\ref{a2}).
  
\section{Local RG equation from deformed AdS space}

\setcounter{equation}{0}

For the general case $\beta^i \neq 0$, we consider a
$(d+1)$-dimensional space with metric
\begin{equation} \label{41}
ds^2 = e^{2A(r)} g_{\mu\nu}(x) dx^{\mu} dx^{\nu}+dr^2\quad.
\end{equation}
Our analysis applies to flows 
from an UV to an IR fixed point, ie. the space is
asymptotically AdS both for $r\to \infty$ and for $r \to
-\infty$.  The bosonic part of the supergravity action is given by 
\begin{equation}\label{action}
  S = \, \frac{1}{16 \pi G_N}\, \left[ \,  \int_{{\cal M}}
 d^{d+1}x \sqrt{G}\left[ { R} + 
    G^{\mu\nu}L^{ij}(\Phi)\partial_{\mu}\Phi_i \partial_{\nu }\Phi_j 
+V(\Phi)\right]\quad - 
\int_{\pr {\cal M}} \! \d^d x \, \sqrt{\gamma} \, 2 \, K \, \right] \; . 
\end{equation}
Here $G$ denotes the $(d+1)$-dimensional 
metric, $\gamma$ is the induced metric on the boundary
 and $L^{ij}$ is the metric on the 
space of the scalar supergravity fields which is positive definite. As may
be derived using either supersymmetry \cite{Freedmanetal} 
or the Hamilton-Jacobi
approach to holographic RG flows \cite{Deboer}, the RG flow is generated
by the first-order equations
\begin{equation}\label{first}
\frac{dA(r)}{dr} = -\frac{g}{3} {\cal W}(\Phi)\;,\frac{\d
\Phi^i(r)}{\d
  r} = \frac{g}{2} L^{ij} \frac{\partial {\cal W}(\Phi)}{\partial\Phi^j}\quad,
\end{equation}
with the superpotential
\begin{equation}
V(\phi) = \frac{g^2}{8} L^{ij}\frac{\partial
  {\cal W}}{\partial\Phi^i}\frac{\partial {\cal W}}{\partial\Phi^j} -
\frac{g^2}{3}{\cal W}(\Phi)^2\quad .
\end{equation}
Here $g$ is the Yang-Mills coupling.

For the definition of the renormalization scale and of the $\beta$
functions we follow \cite{Behrndt,Anselmi,Porrati1} and define
\begin{equation}
  \label{beta}
  \ln\mu \equiv A(r)\, , \quad \beta^i \equiv
  \mu\frac{\d}{\d\mu}\Phi^i =  \frac{d \Phi^i}{dA(r)}\quad.
\end{equation}
Using ({\ref{first}}) we have
\begin{equation}\label{betaW}
\beta^i = \, -
\frac{3}{2}\frac{1}{\W} L^{ij}\frac{\partial \W}{\partial\Phi^j} \, .
\end{equation}
The choice (\ref{beta}) for the
definition of the $\beta$ functions corresponds to the choice of a
particular renormalization scheme. However requiring (\ref{betaW}) to be
covariant under a change of scheme puts some restrictions on the possible
definitions for $\beta^i$ : Suppose we would define 
\begin{equation}
  \beta^i \equiv  \frac{\d \Phi^i}{\d f(r)}
\end{equation}
with some arbitrary function $f(r)$. Then
\begin{equation}
  \beta^i =  \frac{\d \Phi^i}{\d f(r)} =   \frac{\d \Phi^i}{\d
    r}\cdot\frac{1}{f'} = \frac{g}{2}L^{ij}\frac{\partial
    \W}{\partial\Phi^j}\cdot \frac{1}{f'} = -\frac{3}{2}
\frac{L^{ij}}{\W}\frac{\partial \W}{\partial\Phi^j}\cdot\frac{A'}{f'}\,.
\end{equation}
We see that covariance of (\ref{betaW}) requires $f$ to be related to $A$
in general, though it is conceivable that for a given theory, redefinitions 
of $\phi_i$ and $\W$ exist such as to ensure the relation (\ref{betaW}) for 
more general $f'$s.

We assume that the PBH transformation as discussed
in section 3 may be generalized to deformed AdS spaces, i.e.~we assume
that there is a $(d+1)$-dimensional diffeomorphism under which both the 
metric (\ref{41}) - after a suitable coordinate transformation to a
gene\-ralized Fefferman-Graham form - and the action $S$ are invariant, 
and which reduces to a Weyl transformation on the $d$-dimensional
hypersurface at $r=r_0$.  $A(r_0)= \ln \mu_0$ defines a renormalization
scale $\mu_0$. For deformed AdS spaces, we expect the supergravity scalars
to acquire an anomalous Weyl weight related to their $\beta$ function.
Of course it would be very interesting to work out the exact form of the
corresponding $(d+1)$-dimensional diffeomorphism. We leave this for future
work. For our purposes here, we assume that the
Weyl transformation induced by the PBH transformation
is of the form
\begin{equation}
\delta \phi_0{}^i (x, r_0) = \, - \, \sigma 
(\Delta^{(i)} -d + \beta^{(i)}) \phi_0{}^i (x,r_0)
\end{equation}
for the lowest order term in the expansion of $\phi^i(r,x)$ in 
$\eps = (r- r_0)^2$.
$\beta^{(i)}$ is related to the $\beta$
function by
\begin{equation}
\beta^i = \beta^{(i)} \Phi^i \, .
\end{equation}
In the limit $\eps \rightarrow 0$, subject to suitable
regularization, the equation expressing $(d+1)$-dimensional diffeomorphism
invariance of the action $S$ then reduces to
\begin{gather}
\lim_{\eps\rightarrow 0} \int \! \d^d x \, \sigma(x) \left(
\, - \, 2 \eps \frac{\delta}{\delta \eps} - 2 g_0{}^{\mu \nu}
\frac{\delta}{
\delta g_0{}^{\mu \nu}} + (\Delta^{(i)} -d + \hat \beta^{(i)}) \phi_0{}^i
\frac{\delta} {\delta \phi_0{}^i} \right) {\rm e}^{S-S_0} \hspace{2.7cm} 
\nonumber\\ 
\hspace{8cm}=
\int \! \d^d x\, \left( \sigma(x) \B + \pr^\mu \sigma Z_\mu \right)
\, , \label{lRG}
\end{gather}
where
\begin{gather}
\int \! \d^d x\,\left( \sigma(x) \B + \pr^\mu \sigma Z_\mu \right)
\hspace{10cm} \nonumber\\ \hspace{3.3cm}
=  \lim_{\eps\rightarrow 0} \, \int \! \d^d x\, \sigma \,\left(
 2 \eps \frac{\delta}{\delta \eps} +  2 g_0{}^{\mu \nu} \frac{\delta}{
\delta g_0{}^{\mu \nu}} - (\Delta^{(i)} -d + \hat \beta^{(i)}) \phi_0{}^i
\frac{\delta} {\delta \phi_0{}^i} \right) S_0  \, . \label{lRGf}
\end{gather}
(\ref{lRGf}) imposes a finiteness condition on the Weyl variation
of the divergent local counterterms $S_0$ since $\B$, $Z_\mu$ have
to be finite. We investigate the implications of this finiteness
condition in detail in section 5 below. Note also that
\begin{equation}
\hat \beta^i = 
 \frac{\d \Phi^i(r,x)}{\d A(r)} = 
 \frac{\d \Phi^i(r_0,x)}{\d A(r_0)} + \O(\eps) = \beta^i + \O(\eps) \, .
\end{equation}
Since the limit in (\ref{lRG}) is finite, we obtain
\begin{equation}
 \int \! \d^d x \, \sigma(x) \left(
 - 2 g_0{}^{\mu \nu} \frac{\delta}{
\delta g_0{}^{\mu \nu}} + (\Delta^{(i)} -d +  \beta^{(i)}) \phi_0{}^i
\frac{\delta} {\delta \phi_0{}^i} \right) {\rm e}^{S-S_0} =
\int \! \d^d x\, \left( \sigma(x) \B + \pr^\mu \sigma Z_\mu \right)
\, . \label{lRG0}
\end{equation}
Since $\sigma(x)$ is an arbitrary function,
this equation is identical to the local Callan-Symanzik equation
obtained in the Hamilton-Jacobi formalism in \cite{Deboer}, up to terms
involving $(\Delta^{(i)} - d) \phi^i{}_{(0)}$. These terms are due to the
fact that (\ref{lRG0}) is a RG equation rather than a Callan-Symanzik
equation\footnote{Note that in standard
quantum field theory,
the RG equation determines how a renormalized theory evolves under
a change of the renormalization scale, while the Callan-Symanzik equation
determines how a renormalized theory evolves under a scale transformation.
The two equations coincide only in the case when the theory is massless
in the classical limit, or more generally when all couplings are 
dimensionless, which corresponds to $\Delta^{(i)} = d$.}.
In the special case when $\Delta^{(i)} = d$
for all $i$, the local Callan-Symanzik equation and (\ref{lRG0})
coincide to give 
\begin{gather}
\label{lRG4}
\int \! \d^d x \, \sqrt{g} \sigma(x) \left( - 2 g_0{}^{\mu \nu} \frac{\delta}
{\delta g_0{}^{\mu \nu}} + \beta^i \frac{\delta}{\delta \phi_0{}^i} \right)
{\rm e}^{S-S_0}
  = \int\!\! \d^4 x \, 
\sqrt{g} \left[ \sigma {\cal B} + \pr^\mu \sigma Z_\mu \right] \,  .
\end{gather}
For general $\Delta^{(i)}$, using that
\begin{gather}
\left(\mu \frac{\pr}{\pr  \mu}\,  + \int \! \d^d x \, \sqrt{g}  
\left( - 2 g_0{}^{\mu \nu} \frac{\delta}
{\delta g_0{}^{\mu \nu}} + (\Delta^{(i)} -d) \phi_0{}^i \frac{\delta}
{\delta \phi_0{}^i} \right) \right) {\rm e}^{S-S_0} = 0 \, ,
\end{gather}
and in the limit when the $x$-dependent fluctuations of $\phi_0{}^i(x,r_0)$
vanish, (\ref{lRG0}) gives
\begin{equation}
(\mu \frac{\pr}{\pr \mu} + \beta^i \pr_i ) {\rm e}^{S-S_0 } = \int \! \d^d x \,
\B \, , \label{gCS}
\end{equation}
for $\sigma(x)= {\rm const.}$.
(\ref{gCS}) is the well-known global form for the Callan-Symanzik equation.

The main point of our analysis is that (\ref{lRG4}) is exactly of the same form
as the field-theoretical local RG equation (\ref{RG}) presented in
section 2.
As a consequence we find that the field-theoretical discussion of
the implications of Weyl consistency on $\B$ and $Z_\mu$ applies in particular
to the holographic RG, such that we may use the results of the field theory 
discussion in order to obtain new information about the holographic flow.
On the field theory side, a basis  for the local anomaly contributions
$\B$ and $Z_\mu$ is given in (\ref{ano4}). On the supergravity side, the
local anomaly contributions were calculated for specific
cases for instance in \cite{Fukuma,Odintsov,Mueck}. These results agree with
the field-theoretical basis.

For analyzing the implications
of the Weyl consistency conditions for the holographic RG flows, we
restrict ourselves first to the case when all active supergravity scalars
are dual to operators with $\Delta^{(i)}=d$, such that (\ref{lRG4})
applies. 

The implications of the Weyl consistency conditions enable us in particular
to relate the $C$ function of supergravity \cite{Freedmanetal,Girardello},
\begin{equation} C(r) \, = \, \frac{c_0}{|\W|^{d-1}} \, , \qquad
\, ,\label{CADS} \end{equation}
whose flow is positive definite due to the weak energy condition, 
to the coefficients in the anomalies $\B$ and
$Z_\mu$. These coefficients are functions of the
scalar supergravity fields once conformal symmetry is broken. Let us first
consider the case $d=2$, for which we know within quantum field theory 
that the $C$ theorem holds. Weyl consistency implies 
the relation
\begin{equation}
\pr_i a  = \chi_{ij} \beta^j - \beta^j \pr_j w_i 
- \pr_i \beta^j \, w_j \, ,
\end{equation}
with the notation of (\ref{RG2}),
For the holographic flows this relation is satisfied in particular by
\begin{gather} 
a = C - L_{ij} \beta^i \beta^j\, , \quad C= \frac{c_0}{|\W|} \, , 
\quad
w_i = L_{ij} \beta^j = \, - \, {\ts \frac{3}{2}} \pr_i ln \W\, , \quad
\chi_{ij} = {\ts \frac{2}{3}}\, C L_{ij} \, .
\label{sol} 
\end{gather}
Of course, this is just a particular solution. In section 5
below, we calculate these anomaly coefficients
in a well-defined regularization scheme.

The result (\ref{sol}) corresponds just to the ``holographic'' scheme
as defined in \cite{Anselmi}. With (\ref{sol}) the equation
\begin{equation} \label{ftflow}
\beta^i \pr_i \tilde{a} = \chi_{ij} \beta^i \beta^j \, , \quad \tilde{a} =
a + \beta^i w_i \, , 
\end{equation}
for the anomaly coefficients corresponds just to the well-known holographic
$C$ theorem
\begin{equation}
\dot{C} = - \beta^i \pr_i C = - {\ts \frac{2}{3}}\, C \,  
 L_{ij} \beta^i \beta^j \, \leq 0 \, ,
\end{equation}
where the inequality follows from the weak energy condition, or
equivalently from the positivity of $L_{ij}\beta^i \beta^j$ and of $C$.
Of course, the results (\ref{sol}) for the anomaly coefficients are scheme
dependent and in particular the addition of finite local counterterms
would change the result in exactly the same way as
obtained in field theory, (\ref{sc}).
This implies that the holographic $C$ theorem is scheme-dependent. For 
instance, $w_i$ may be set to zero, and correspondingly
$a$ to $C$,  by adding a finite local term
\begin{equation}
S_f{}^R = \, - \, {\ts \frac{3}{2}}\, \int \! \d^2x \sqrt{g} \, (\ln \W) \, R \, 
\end{equation}
to the vacuum energy functional. $\tilde{a} =
a + \beta^i w_i$ remains invariant under this change. 
Moreover, when a finite local boundary term 
\begin{equation} S_f{}^\pr = \, 
\int \! \d^2x \sqrt{g} \, c_{ij} \pr^\mu \phi_0{}^i \pr_\mu \phi_0{}^j 
\end{equation}
is added to the vacuum
energy functional such that
\begin{equation}
\delta \tilde{a} = c_{ij} \beta^i \beta^j \, , \quad \delta \chi_{ij} = 
\L_\beta c_{ij} \, ,
\end{equation}
with $\L_\beta$ the Lie derivative as in (\ref{Lie}),
then although the equation (\ref{ftflow}) remains invariant under this 
change of scheme, it is not clear if  the new
\begin{equation}
\chi'_{ij} \equiv \chi_{ij} + \, \delta \chi_{ij} = \, {\ts \frac{2}{3}} \, 
C \, L_{ij} + \L_\beta c_{ij}
\end{equation}
remains positive.
In two dimensions however, the field theory $C$ theorem holds independently
of the scheme chosen. As discussed in section 2, within the Weyl consistency
approach, the key ingredient for the proof of the C theorem is 
$
G_{ij} = \chi_{ij} + \L_\beta c_{ij} \, 
$
with $G_{ij}$ the Zamolodchikov metric. This implies that
$\chi_{ij} + \L_\beta c_{ij} \, $ is positive definite independently of
the scheme, such that the flow of
$
C_{\rm ft} \equiv \tilde{a} + c_{ij} \beta^i \beta^j
$
is positive for any renormalization scheme, ie.~independently of the explicit 
form of $c_{ij}$. For the holographic flows in $d=2$ this implies that the 
$C$ theorem holds independently of the form of possible finite  
counterterms. We note that for the ``holographic'' scheme, which 
corresponds to $c_{ij}=0$, ie.~to minimal subtraction, the field-theoretical
and the holographic $C$ theorem coincide.

Similarly for $d=4$ we may show that for the holographic scheme, the 
holographic C function (\ref{CADS}), which satisfies
\begin{equation} \label{flow4}
-\beta^i \pr_i C = \, - \, 2 \,C \, L_{ij} \beta^i \beta^j \, ,
\end{equation}
coincides with the field theory relation
\begin{equation} \label{sol4}
\beta^i \pr_i \tilde{a} = {\ts \frac{1}{8}}
\chi^g_{ij} \beta^i \beta^j \, , \quad \tilde{a} = a + \, \achtel 
\beta^i w_i \, , 
\end{equation}
with the anomaly coefficients defined by (\ref{ano4}), if 
\begin{gather}
a = C  - L_{ij} \beta^i \beta^j 
\, , \quad w_i = 8 \,  L_{ij} \beta^j \, \quad \chi^g_{ij} = 16 \, 
C \, L_{ij} \, , \quad \tilde{a} = C \, ,
\end{gather}
or equivalently - after  adding the appropriate finite counterterm - 
by
\begin{gather} \label{consistent}
a = C \, , \quad w_i = 0 \, \quad \chi^g_{ij} = \, 16 \, C \, L_{ij} \, .
\end{gather}
Again the relation (\ref{sol4}) is invariant under the 
addition of finite boundary counterterms under which $\chi^g_{ij} \rightarrow
\chi^g_{ij} + \L_\beta c_{ij}$.  However in
$d=4$ it has not yet been possible to relate $\chi^g_{ij}+ \L_\beta c_{ij}$
to the Zamolodchikov metric, such as to show positivity. 
Therefore, from
a field-theoretical point of view, the holographic C theorem in $d=4$
is a scheme-dependent result.

\section{Anomaly coefficients within minimal subtraction}

\setcounter{equation}{0}

In this section we show that the particular solution
for the anomaly coefficients which corresponds to the holographic C theorem
(\ref{flow4}), with $w^i$ set to zero as in (\ref{consistent}), 
may be obtained as a
consistent solution of equation (\ref{lRGf}), which imposes a finiteness
condition on the Weyl variation of the divergent local counterterms
$S_0$\footnote{Within standard 
perturbative quantum field theory, finiteness
conditions of the form (\ref{lRGf}) have been considered in \cite{Jack}.}.
For definiteness we consider the 
four-dimensional case, the argument for $d=2$ being exactly analogous. 
As before, we restrict the calculation to the case that all active scalars
have $\Delta^{(i)} = d$.
We solve (\ref{lRG4}) order by order in a perturbative approach, 
expanding around  $r=r_0$ and in derivatives with respect to the couplings.
For instance for the superpotential we have the 
expansion
\begin{align} \label{potentialexpansion}
\W(\Phi(r)) &= \W(\Phi(r_0)) + \pr_i \W(\Phi(r_0)) \delta \Phi^i(\eps)
+ \half \pr_i \pr_j \W(\Phi(r_0)) \delta \Phi^i(\eps) \delta \Phi^j(\eps)
+ \dots \, , \\  \, \Phi^i(r) &= \phi_0{}^i(r_0) + \eps \phi_2{}^i\, (r_0)
+ \dots \, \\
\delta \Phi^i(\eps)
&= \delta \phi_0{}^i(r_0) + \eps \delta \phi_2{}^i (r_0) 
+ \dots \, ,  \quad \eps =  (r-r_0)^2  \, .
\end{align}
Here $\eps$ is the regulator.
According to \cite{Skenderis}, we expect logarithms of $\eps$ 
in the expansion
of $\delta \Phi^i(\eps)$. $\phi_2$ and all higher order contributions
may be expressed in terms of $\phi_0$, but their explicit form
is not rele\-vant here.
Note that due to (\ref{betaW}), the beta functions correspond to one 
derivative of the superpotential. Due to (\ref{first}) we expect
this expansion to be equivalent to a Taylor expansion of $A'(r)$. In the
conformal case, $A''$ and all higher order derivatives vanish.

For solving (\ref{lRGf}) perturbatively for $d=4$, we note that
a basis for the anomaly terms $\B$, $Z_\mu$ 
is given by (\ref{ano4}). Here we
restrict ourselves to considering only those contributions to the anomaly
which are relevant for the C theorem, ie.~those involving the
coefficients $a$, $w_i$ and $\chi^g{}_{ij}$. The divergent local
counterterm contributing to these anomaly contributions is given by
\begin{gather}
S_0{}^C = \, - \, 
\int \! \d^4 x \, \sqrt{g} \left[ b \tilde R \tilde R
+ \, 
{\half} \, \eta_{ij} G^{\mu \nu} \pr_\mu \phi_0{}^i
\pr_\nu \phi_0{}^j  \right]
\end{gather}
with divergent coefficients $b$, $\eta_{ij}$.
Here $\tilde{R} \tilde{R}$ is the Euler density and $G^{\mu \nu}$ is
the Einstein tensor. The finiteness condition  (\ref{lRGf})
implies
\begin{align}
(2 \eps \pr_\eps - \hat \beta^i \pr_i ) \, b \, &= \, a \nonumber\\
(2 \eps \pr_\eps - \L_\beta ) \, \eta_{ij} \, &= \, \chi^g{}_{ij}
\label{set} \\
8 \pr_i b \, - \, \eta_{ij} \hat \beta^j \, &= \, w_i \, . \nonumber 
\end{align}
$\L_\beta \eta_{ij} \equiv \beta^k \pr_k \eta_{ij} + \pr_i \beta^k \eta_{kj}
+ \pr_j \beta^k \eta_{ik}$ is the Lie derivative.
The right hand side of these equations - and therefore also the left hand
side - must be finite. To lowest order, ie.~up to terms involving 
derivatives of $\W(\phi)$ and $L_{ij}(\phi)$, (\ref{set}) is solved by
\begin{align}
b \, &= \,  \half  \ln \eps \, \frac{c_0}{| \W|^3} \equiv \half \ln \eps \, C
 \, , \quad a \, = \, \frac{c_0}{| \W|^3} \equiv C \, \nonumber\\
\eta_{ij} \, &= 8 \, \ln \eps \, C \, L_{ij} \, , \quad \chi_{ij} = 16 \, C \, 
L_{ij} \label{setsol0} \\
w_i &= 0 \, . \nonumber
\end{align}
We see that there is agreement with the result $\beta^i \pr_i \tilde{a}
 = {\ts \frac{1}{8}} \, \chi^g{}_{ij} \phi_0{}^i \phi_0{}^j{}$
obtained from
the Weyl consistency condition. To next order, ie.~including
 terms involving
two derivatives with respect to $\phi^i$, (\ref{set}) has the
solution
\begin{align}
b \, &= \,   \half \ln \eps \, C \, + {\ts \frac{1}{4}} \ln^2 \eps \,
C \, L_{ij} \beta^i \beta^j 
 \, , \quad a \, = \,  C \, , \nonumber\\
\eta_{ij} \, &= 8 \, \ln \eps \, C \, L_{ij} \, + 2 \ln^2 \eps \,
 \L_\beta ( C L_{ij}) 
\, , \quad \chi_{ij} = 16 \, C \, 
L_{ij} \, ,  \label{setsol2} \\
w_i &= \ln \eps ( 4 \pr_i C - 8 C\, L_{ij} \beta^j ) = 0 \, . \nonumber
\end{align}
The contributions to $a,\chi,w$ involving $\ln \eps$ cancel and the 
contributions involving $\ln^2 \eps$
are of higher order since they contain at least three
 derivatives with respect to 
$\phi^i$.
Therefore we have obtained the finite result
\begin{gather}
a=C, \quad \chi_{ij} = 16 \, C \, L_{ij}, \quad w_i=0 \label{anoresult} 
\end{gather}
 satisfying the Weyl consistency
condition to second order in the derivatives with respect to $\phi^i$.
The appearance of $\ln^2 \eps$ is natural in a perturbation expansion
and the essential feature is that the result for the anomalies is finite
to a given order in the expansion. To obtain the result (\ref{anoresult})
including terms involving four
derivatives, we have to include terms involving 
$\ln^3 \eps$ in order to cancel the terms involving $\ln^2 \eps$ :
\begin{align}
b \, &= \,   \half \ln \eps \, C \, + {\ts \frac{1}{4}} \ln^2 \eps \,
C \, L_{ij} \beta^i \beta^j\, +\, {\ts \frac{1}{24}} \ln^3 \eps \,
 \beta^k \pr_k
(C L_{ij} \beta^i \beta^j) 
 \, , \quad a \, = \,  C \, , \nonumber\\
\eta_{ij} \, &= 8 \, \ln \eps \, C \, L_{ij} \, + 2 \ln^2 \eps
 \, \L_\beta ( C L_{ij}) \, + \, {\ts \frac{1}{3}} \, \ln^3 \eps \, 
\L_\beta ( \L_\beta ( C L_{ij}))  \, , \quad \chi_{ij} = 16 \, C \, 
L_{ij} \, , \label{setsol3} \\
w_i &= \ln^2 \eps ( 2 \, \pr_i ( C L_{jk} \beta^j \beta^k)
- 2 \L_\beta (C L_{ij}) \beta^j ) = 0 \, . \label{setsolw}
\end{align}
The calculation showing that $w_i$ in (\ref{setsolw}) vanishes is non-trivial
and relies on the result (\ref{beta}) expressing $\beta^i$ in terms
of the superpotential and its derivative. We expect that we may continue
this calculation to any given order in the derivatives of $\W$. It
seems possible to use induction to show that the result (\ref{anoresult})
holds to all orders in the expansion. 

These results show that for operators with
$\Delta^{(i)} = d$, the holographic C theorem may be obtained
from a consistent perturbative solution of the finiteness condition
(\ref{lRGf}) in a well-defined regularization scheme, together with the
Weyl consistency condition originating from the
local structure of the holographic RG.


\section{Conclusion and Perspectives}

\setcounter{equation}{0}

For our analysis of holographic RG flows
we have made use of results for local RG equations
obtained within standard quantum field theory. 
There are further examples of field theory
results relevant to holographic flows, for instance concerning the possibility
of constructing a four-dimensional C function based on the energy-momentum 
tensor two-point function. This two-point function is generally of the form
\begin{align} \label{om}
\l T_{\mu \nu}(x) T_{\si \rho}(0) \r \,  & = \, 
{\ts \frac{1}{3}} \, S_{\mu \nu} S_{\si \rho} \, \Omega_0(x)
+ ( S_{\mu(\si} S_{\rho)\nu} - {\ts \frac{1}{3}} S_{\mu \nu} S_{\si \rho}
) \, \Omega_2(x) \, , \\  S_{\mu \nu} & \equiv \delta_{\mu \nu} \pr^2
- \pr_\mu \pr_\nu \, . \nonumber
\end{align}
It was shown in \cite{Shore}, based on the spectral 
representation approach to the C theorem
\cite{spectral},  that in four dimensions the function $C_F(\lambda)$,
which is directly related to $\Omega_2$ in (\ref{om}) and which 
at the fixed points coincides with 
the coefficient $c$  of the Weyl tensor squared contribution to the
anomaly, satisfies
\begin{gather}
- \beta^i \pr_i \, C_F \, = \, 2 \, C_F \, - G \, , \label{G}
\end{gather}
with $G$ positive definite\footnote{For a 
precise definition of $C_F$ and $G$ see
\cite{Shore,Osborn}.}. This result indicates that in general the flow of 
$C_F$ is not expected to be monotonic. In fact, counterexamples
for a C theorem based on $C_F$ have been found in \cite{Grisaru}. As discussed
for instance in section 2 above, it is
expected that a four-dimensional C theorem will involve a C function 
which at the fixed points coincides with the coefficient $a$ of the Euler
density anomaly (see also \cite{OF}). 
However for holographic flows, whose anomaly coefficients
generally satisfy $a=c$ and which therefore represent a special class
of field theories, examples
for which  the flow of $C_F$ is indeed monotonic have been
discussed in \cite{Porrati2}. It will therefore 
be interesting to determine the
exact form of (\ref{G}) for holographic flows.
- Moreover the field-theoretical analysis of the geometrical structure of the
renormali\-zation group  in \cite{Dolan,Kar}
is also related to the holographic RG.

Furthermore it will be 
very interesting to investigate if the results of this paper
may be applied to examples for holographic flows which go beyond
the supergravity approximation. For example, the structure of
the conformal anomaly
in the Polchinski-Strassler model \cite{Strassler}
has been investigated very recently in \cite{Taylor}. We expect that
the local RG  may be useful for obtaining further results
about the conformal anomaly in this and in related models.

In \cite{Gimon}, the question was raised if it is possible to find a
general criterion for the existence of a holographic dual for any
given quantum field theory. In \cite{Kehagias} it was suggested for
four-dimensional 
conformal field theories that a necessary criterion for the existence
of a gravity dual is that the coefficients of the Euler and Weyl tensor
squared anomalies must coincide, $a=c$. In the light of our results
here, and noting that in the framework of perturbative
quantum field theory, local RG equations have been shown to hold for a
number of renormalizable quantum field theories, including $\phi^4$-theory
and gauge theories in four dimensions, we may speculate that it
may be possible to construct holographic duals also for theories
with $a \neq c$, starting from their local RG. These gravity duals, if
they exist, would presumably be of a very different form than those
obtained from deformations of the AdS/CFT correspondence.

\vspace{2cm}

{\bf Acknowledgements}

I am grateful to Djordje Minic and to Dan Freedman for very useful
discussions. Moreover I would like to thank the CIT-USC Center for
Theoretical Physics for hospitality during the course of this work.

This work was supported by an Emmy Noether fellowship of the
Deutsche Forschungsgemeinschaft (DFG).

\end{document}